\documentclass[fleqn,usenatbib,useAMS]{mnras}

\usepackage{graphicx}	
\usepackage{amsmath}
\usepackage{amssymb}	
\usepackage{newtxtext,newtxmath}
\usepackage{physics} 
\usepackage[T1]{fontenc}

\newcommand{\p}{\partial}
\newcommand{\OmK}{\Omega_\text{K}}

\newcommand{\nn}{\nonumber}
\newcommand\bb[1]{\mbox{\boldmath{$#1$}}}

\begin{document}
\title[]{Rossby wave instability in magnetized protoplanetary disks. I. azimuthal and vertical B-fields}

\author[]{Can Cui$^{1,2}$\thanks{\href{mailto:can.cui@astro.utoronto.ca
}{can.cui@astro.utoronto.ca
}}, Ashutosh Tripathi$^{2}$, Cong Yu$^{3,4}$, Min-Kai Lin$^{5,6}$ and Andrew Youdin$^{7,8}$ 
\\ 
\\ \\ 
$^{1}$Department of Astronomy and Astrophysics, University of Toronto, Toronto, ON M5S 3H4, Canada \\
$^{2}$DAMTP, University of Cambridge, Wilberforce Road, Cambridge CB3 0WA, UK \\
$^{3}$School of Physics and Astronomy, Sun Yat-Sen University, Zhuhai 519082, China\\
$^{4}$CSST Science Center for the Guangdong-Hong Kong-Macau Greater Bay Area, Zhuhai 519082, China\\
$^{5}$Institute of Astronomy and Astrophysics, Academia Sinica,Taipei 10617, Taiwan, R.O.C. \\ 
$^{6}$Physics Division, National Center for Theoretical Sciences, Taipei 10617, Taiwan, R.O.C. \\ 
$^{7}$Department of Astronomy and Steward Observatory, University of Arizona, Tucson, AZ 85721, USA \\
$^{8}$The Lunar and Planetary Laboratory, University of Arizona, Tucson, AZ 85721, USA \\
}

\pubyear{2024}

\label{firstpage}
\pagerange{\pageref{firstpage}--\pageref{lastpage}}
\maketitle

\begin{abstract}

Rossby wave instability (RWI) is considered the underlying mechanism to crescent-shaped azimuthal asymmetries, discovered in (sub-)millimeter dust continuum of many protoplanetary disks. Previous works on linear theory were conducted in the hydrodynamic limit. Nevertheless, protoplanetary disks are likely magnetized and weakly ionized. We examine the influence of magnetic fields and non-ideal magnetohydrodynamic (MHD) effects - namely, Ohmic resistivity, Hall drift, and ambipolar diffusion - on the RWI unstable modes. We perform radially global linear analyses, employing constant azimuthal ($B_\phi$) or vertical ($B_z$) background magnetic fields. It is found that, in the ideal MHD regime, magnetism can either enhance or diminish RWI growth. Strong non-ideal MHD effects cause RWI growth rates to recover hydrodynamic results. The sign of Hall Els\"{a}sser number slightly complicates the results. Vertical wavenumbers can diminish growth rates. 

\end{abstract}


\section{Introduction}\label{in}

Protoplanetary disks are composed of gas and dust orbiting pre-main-sequence stars \citep{armitage11}. They are the birth place of planets, where micron-sized dust grains coalesce and evolve into km-sized planetesimals, eventually giving rise to terrestrial planets or gas giants cores. Nevertheless, the growth of dust grains faces several barriers, including bouncing, fragmentation, and fast radial drift \citep{weidenschilling77,guttler+10,zsom+10}.
Rossby wave instability (RWI) is perhaps a promising mechanism to circumvent these barriers. Its non-linear state generates large, lopsided crescent-shaped vortices, which concentrate grains towards pressure maximum, fostering streaming instability and subsequent gravitational collapse \citep{gw73,yg05}. These vortices manifest as azimuthal asymmetries observed in (sub-)millimeter dust continuum as well as CO rotational transition lines \citep[e.g.,][]{huang_etal18,vdmarel+21}. Notable examples include IRS 48 \citep{vdmarel_etal13}, HD142527 \citep{muto+15}, and AB Aur \citep{fuente+17}. 

The necessary condition of RWI is local extrema in the radial vortensity profile, $(\Sigma\Omega/\kappa^2)S^{2/\Gamma}$ \citep{lovelace99,li_etal00,chang23,chang24}.
It gives exponential growth of non-axisymmetric modes ($\propto\exp[im\phi]$, where $m=1,2,...$) on each side of the corotation radius. These unstable Rossby modes are confined between the inner and outer Lindblad resonances, where density waves are launched and propagate away \citep{shu+64,gt79}.
The unstable Rossby waves manifest as anticyclonic vortices, characterized by vorticity ($\nabla \times \delta \mathbf{u}$) directed oppositely to the rotation of the disk, yielding a maximum perturbed pressure at the vortex core \citep{lovelace+14}. Extensions of the classic linear theory incorporating self-gravity \citep{lp11,lin12}, magnetic fields \citep{yu09,yu13}, thermal relaxation \citep{hy22}, and aerodynamic drag between dust and gas \citep{lb23} have been explored to obtain a more complete understanding of the RWI.

In the context of protoplanetary disks, numerical simulations of the RWI have been commonly performed at the gap edges carved by a planet \citep[e.g.,][]{valvorro+07,zhu+14,zs14,hammer+17,huang+18,li+20,cimerman+23}, or at the dead zone edges of the magneto-rotational instability \citep[MRI; e.g.,][]{lyra12,miranda+16,miranda+17}, for which local vortensity extrema take place. These simulations elucidate that the non-linear saturation of the Rossby vortices is primarily governed by $m=1$ modes \citep{godon99,li+01,meheut+12}. Furthermore, the long-term survival of Rossby vortices are examined under secondary instabilities \citep[e.g.,][]{LP09}, thermal relaxation \citep[e.g.,][]{fung+21,rometsch+21}, dust-gas two-fluid framework \citep[e.g.,][]{surville+16,surville+19,li+20}, and self-gravity \citep{lyra+24}.

While the RWI is inherently a hydrodynamic phenomenon, it is crucial to acknowledge the magnetized nature of protoplanetary disks, where magnetic fields likely originate from the primordial molecular clouds \citep{galli93,girart+06,girart+09}. Owing to the weak thermal ionization of passively heated protoplanetary disks and the modest non-thermal ionization by, for instance, the stellar FUV, EUV, X-rays, and cosmic rays \citep{wardle07,bai11a}, the gas and magnetic fields are not perfectly coupled. The disk material may be weakly ionized, and three non-ideal MHD effects come into play: Ohmic resistivity, the Hall drift, and ambipolar diffusion \citep{wbf21,lesur21}. 

At the same ionization fraction, Ohmic, Hall, and ambipolar diffusivities are respectively proportional to $\eta_\mathrm{O}\propto \mathrm{const.}$, $\eta_\mathrm{H}\propto B/\rho$, and $\eta_\mathrm{A}\propto B^2/\rho^2$ \citep[e.g.,][]{bai17}. Consequently, Ohmic resistivity dominates in high-density regions near the inner disk and at the midplane, while ambipolar diffusion becomes prominent in the outer disk. Hall drift is most effective between these regions. Both numerical simulations and analytical theory have underscored the significance of non-ideal MHD effects in shaping disk structure and evolution, such as dead zone structures \citep[e.g.,][]{flock+17rad}, magnetized wind kinematics \citep[e.g.,][]{bs13,bai+16,wang_etal19}, dust dynamics and distribution \citep[e.g.,][]{lin+22,xb22,wu+24}, heating and disk temperature profiles \citep[e.g.,][]{shoji+19,bl20}, and the formation of annular substructures \citep[e.g.,][]{rl19,cb21,cb22}.

Previously, the non-ideal MHD effects were incorporated in the linear studies of several instabilities applicable to protoplanetary disks, for example, the MRI \citep{bh91} and the vertical shear instability \citep{gs67,nelson_etal13}. These investigations unveiled significant alterations in the unstable modes induced by non-ideal MHD physics. In the context of MRI, both Ohmic resistivity and ambipolar diffusion tend to stabilize the instability, although under specific conditions, ambipolar diffusion shear instability can be excited \citep{bb94,kb04,lk22}. The impact of Hall drift is twofold, based upon the sign of the Hall Els\"{a}sser number \citep{wardle99,wardle12}. Concerning the vertical shear instability, strong ionization can substantially weaken the unstable modes, while subdued modes regain strength under weak ionizations \citep{lp18,cb20,cl21,lk22}.

To deepen our comprehension of RWI behavior in magnetized disks, investigations must encompass both ideal and non-ideal MHD physics. Previous studies have primarily focused on the ideal MHD regime. \citet{yu09} utilized Lagrangian perturbation theory to probe the influence of toroidal magnetic fields on the RWI. Their findings revealed a continuous reduction in growth rates with increasing magnetization. Subsequently, \citet{yu13} explored the impact of poloidal fields using a vertically integrated disk model. Interestingly, they observed a dichotomy in results: pure vertical fields typically diminish RWI growth rates, whereas the presence of radial fields tends to enhance them. 

In this work, we explore the RWI unstable modes in the ideal and non-ideal MHD limit, employing three-dimensional linear analysis with Eulerian perturbations. The paper is organized as follows. In Section \ref{sec:th}, we introduce the governing dynamical equations and the perturbation equations that delineate our theoretical framework. Section \ref{sec:me} elaborates on the numerical methodologies employed to solve the set of ordinary differential equations (ODEs) governing the magnetized RWI. In Section \ref{sec:re}, we present numerical solutions obtained and elucidate the RWI growth rate behaviors. We summarize the main findings in Section \ref{sec:c}.

\section{Theory}\label{sec:th}
 
\subsection{Dynamical equations}\label{sec:de}

The stability of a three-dimensional, compressible, magnetized disk with background radial vortensity extrema is analyzed in cylindrical coordinates ($r,\phi,z$). The gravitational potential is given by $\Phi=-GM/(r^2+z^2)^{1/2}$, where $M$ is the mass of central star. Disk self-gravity is neglected. The governing equations for this magnetized disk in Gaussian units are the continuity, momentum, and entropy conservations, 
\begin{equation}
\frac{d\rho}{dt}+\rho\nabla\cdot\bb{v}=0,
\label{eq:1}
\end{equation}
\begin{equation}
\frac{d\bb{v}}{dt} + \frac{1}{\rho}\nabla\bigg[P+\frac{B^2}{8\pi}\bigg]+\nabla\Phi - \frac{1}{4\pi\rho}(\bb{B}\cdot\nabla)\bb{B}=0,
\label{eq:2}
\end{equation}
\begin{equation}
\frac{dS}{dt}=0,
\label{eq:3}
\end{equation}
where the material derivative is defined as $d/dt\equiv\p/\p t+v\cdot\nabla$, and $S\equiv P/\rho^\Gamma$ is the entropy of the disk matter. 
The induction equation is written as, 
\begin{equation}
\frac{\p\bb{B}}{\p t}-\nabla\times(\bb{v}\times\bb{B}-c\bb{E}')=0.
\label{eq:ind}
\end{equation}
The non-ideal MHD terms manifest in the electric field of the rest fluid frame,
\begin{equation}
\bb{E}' = \frac{4\pi}{c^2}[\eta_\mathrm{O}\bb{J}+\eta_\mathrm{H}\bb{J}\times\bb{b}-\eta_\mathrm{A}(\bb{J}\times\bb{b})\times\bb{b}], 
\end{equation}
where the unit vector of magnetic field is denoted by $\bb{b}=\bb{B}/|B|$, and Ohmic, Hall, ambipolar diffusivities are denoted by $\eta_\mathrm{O}$, $\eta_\mathrm{H}$, and $\eta_\mathrm{A}$, respectively. The current density is $\bb{J}=c\nabla\times\bb{B}/4\pi$.
Using the divergence free condition $\nabla\cdot\bb{B}=0$, the induction equation can be cast into 
\begin{equation}
\frac{d\bb{B}}{dt} -(\bb{B}\cdot\nabla)\bb{v} +(\nabla\cdot\bb{v})\bb{B} + c\nabla\times\bb{E}'=0.
\end{equation}

\subsection{Equilibrium of the disk}\label{sec:eqm}

The equilibrium disk model is stationary ($\p/\p t=0$), axisymmetric ($\p/\p \phi=0$), and radially global. All background quantities are independent of $z$ ($\p/\p z=0$). The steady-state physical quantities are denoted by the subscript ``0''. The equilibrium velocity field has only the azimuthal component, $v_{\phi0}=\Omega_0 r$. 

The necessary condition for Rossby wave instability is the presence of radial vortensity extrema. This can be achieved by setting up a Gaussian bump centered at $r=r_0$ in the density profile \citep[e.g.,][]{li_etal00},
\begin{equation}
\frac{\rho_\mathrm{0}}{\rho_{00}} = 1+ (A-1)\exp\bigg[-\frac{1}{2}\bigg(\frac{r-r_0}{\Delta r}\bigg)^2\bigg],
\end{equation}
where $\rho_{00}$ is the background density profile without the Gaussian bump. Despite of the radially global nature of the disk model, $\rho_{00}$ is taken to be a constant for simplicity. Setting it to a power-law distribution will not qualitative alter the results as noted in \citet{li_etal00}.

To compute the background pressure $P_0$, we consider a barotropic flow, and hence the pressure is only a function of density $P_0(\rho_0)$, expressed by 
\begin{equation}
\frac{P_0}{P_{0\ast}}= \bigg[\frac{\rho_0}{\rho_{0\ast}}\bigg]^\Gamma,
\end{equation}
where subscript ``$0\ast$'' denotes background quantities evaluated at $r_0$, and $\Gamma$ is the adiabatic index. The adiabatic sound speed is defined as $c_{s0}\equiv(\Gamma P_0/\rho_0)^{1/2}$. 
By specifying $c_{s0\ast}$, we can obtain $P_{0\ast}$ and subsequently $P_0$. Throughout the paper, we set $GM = \rho_0 = r_0=1$, $\Delta r/r_0=0.05$, $\Gamma=5/3$, $A=1.5$, and the disk aspect ratio $c_{s0\ast}/v_{\phi0\ast}=0.06$ at $r_0$. The pressure scale height $H\approx c_{s0\ast}/\Omega_\mathrm{K0\ast}$, where $\Omega_\mathrm{K0\ast}$ is the Keplerian angular speed at $r_0$.

We construct a simplified equilibrium solution for this study, setting $\bb{B}_0$ a constant vector. In equilibrium, the radial momentum equation gives
\begin{equation}
\frac{v_{\phi 0}^2(r)}{r}=\frac{1}{\rho_0}\frac{\p P_0(r)}{\p r} + \frac{\p\Phi(r)}{\p r}.
\end{equation}
In the ideal MHD limit, the $\phi$-component of induction equation is
\begin{equation}
\bigg[B_{r0}\frac{\p}{\p r}+B_{z0}\frac{\p}{\p z}-\frac{B_{r0}}{r}\bigg] v_{\phi 0}=0.  
\label{eq:10}
\end{equation}
It is immediately seen that $B_{\phi0}$ is not involved in these equations, and hence is free to specify. Then for $B_{r0}=0$, we are completely free to specify $B_{z0}$ as $\p/\p z=0$, and hence the constant equilibrium magnetic field is taken to be $\bb{B}_0=(0,B_{\phi0},B_{z0})$. 

When considering non-ideal MHD physics to equilibrium solutions, only the induction equation \eqref{eq:ind} is modified. The current density $\bb{J}$ vanishes for a pure $B_z$ field, and eq \eqref{eq:10} is strictly satisfied. However, the existence of curvature terms from cylindrical coordinates render non-zero $\nabla\times c\bb{E}'$ for a pure $B_\phi$ field. Hence, our equilibrium magnetic field is only valid if we ignore the curvature terms in curl operator. 

The strengths of $B_{\phi0}$ and $B_{z0}$ are parameterized by plasma $\beta$, defined as the ratio of gas pressure to magnetic pressure, 
\begin{equation}
\beta=\frac{8\pi P_0}{B_0^2},
\end{equation}
where $\beta >1$ is commonly satisfied in protoplanetary disks. 
Finally, we quantify the strength of non-ideal MHD effects by their respective Els\"{a}sser numbers,  
\begin{equation}
\Lambda=\frac{v_\textrm{A}^2}{\eta_\mathrm{O}\OmK},
\end{equation}
\begin{equation}
\mathrm{Ha}=\frac{v_\mathrm{A}^2}{\eta_\mathrm{H}\OmK},
\end{equation}
\begin{equation}
{\rm Am}=\frac{v_\mathrm{A}^2}{\eta_\mathrm{A}\OmK},
\end{equation}
where the Alfv\'{e}n velocity is $v^2_{\mathrm A}=B_0^2/4\pi\rho$, and $\OmK$ is the Keplerian angular speed.

\subsection{Perturbations of the disk}\label{sec:pertb}

Consider small perturbations to eqs \eqref{eq:1}-\eqref{eq:ind}, such that $\bb{v} = \bb{v}_0+\delta\bb{v}(r,z,\phi,t)$, $\bb{B} = \bb{B}_0+\delta\bb{B}(r,z,\phi,t)$, ..., . We linearize these equations by considering Eulerian perturbations $\propto f(r)\exp(ik_zz+im\phi-i\omega t)$, where $k_z$ is the vertical wavenumber, $m$ is the azimuthal mode number, and $\omega=\omega_r+i\gamma$ is the mode frequency,  where $\gamma$ denotes the growth rate. We further define the Doppler-shifted wave frequency $\Delta\omega=\omega-m\Omega$, the azimuthal wavenumber $k_\phi=m/r$, and the radial epicyclic frequency $\kappa=[r^{-3}d(r^4\Omega^2)/dr]^{1/2}$. 

We now drop subscript ``$0$'' for the background quantities throughout the rest of the paper. Our model encompasses eight perturbed quantities, $\delta\bb{v},\delta\bb{B},\delta\rho,\delta\Psi$, where
\begin{equation}
\delta\Psi = \frac{\delta P}{\rho},
\label{eq:psi}
\end{equation}
and 
\begin{equation}
\frac{\p\delta\Psi}{\p r} = \frac{1}{\rho}\frac{\p\delta P}{\p r} - \frac{1}{\rho}\frac{\p\rho}{\p r}\delta\Psi.
\end{equation}

We follow \citet{lovelace99}'s and \citet{li_etal00}'s original paper, and define 
radial length scales of entropy, pressure, and density variations as 
\begin{equation}
\mathrm{L_S \equiv \frac{\Gamma}{d \ln S/dr}},
\end{equation}
\begin{equation}
\mathrm{L_P \equiv \frac{\Gamma}{d \ln P/dr}},
\end{equation}
\begin{equation}
\mathrm{L_\rho \equiv \frac{1}{d \ln\rho/dr}}.
\end{equation}
These length scales are related by
\begin{equation}
\mathrm{\frac{1}{L_P} = \frac{1}{L_S} + \frac{1}{L_\rho}}.
\end{equation}
For a barotropic flow considered throughout, the length scale of entropy approaches infinity, $1/\mathrm{L_S}\rightarrow 0$.

To present the perturbation equations optimally, we separate them into two cases: $k_z=0$ (\S\ref{sec:pertb}) and $k_z\neq 0$ (Appendix \ref{app:kz}). 
We first show the set of linearized equations in the ideal MHD limit. In the non-ideal MHD limit, we split the equations into pure vertical and toroidal magnetic field regimes. 

\subsubsection{ideal MHD}

We first derive the linearized equations in the ideal MHD limit. For continuity equation \eqref{eq:1} it is
\begin{equation}
\frac{\p\delta v_r}{\p r} + \bigg[\frac{1}{r}+\frac{1}{\mathrm{L_\rho}}\bigg]\delta v_r + ik_\phi\delta v_\phi - i\Delta\omega \frac{\delta\Psi}{c_s^2}  =0.
\label{eq:17}
\end{equation}
The linearized momentum equations \eqref{eq:2} are
\begin{align}
i\Delta\omega&\delta v_r + 2\Omega\delta v_\phi - \frac{\p\delta\Psi}{\p r} \nn \\ 
&+\frac{1}{4\pi\rho}\bigg[ik_\phi B_\phi\delta B_r-\frac{2B_\phi}{r}\delta B_\phi 
-B_z\frac{\p \delta B_z}{\p r}-B_\phi\frac{\p \delta B_\phi}{\p r}+\frac{B_\phi^2}{r}\frac{\delta\rho}{\rho}\bigg]\nn \\ 
&=0,
\end{align}
\begin{equation}
i\Delta\omega\delta v_\phi - \frac{\kappa^2}{2\Omega}\delta v_r - ik_\phi\delta \Psi + \frac{1}{4\pi\rho}\bigg[\frac{B_\phi}{r}\delta B_r-ik_\phi B_z\delta B_z\bigg] = 0,
\end{equation}
\begin{equation}
i\Delta\omega\delta v_z + \frac{ik_\phi B_\phi}{4\pi\rho}\delta B_z =0.
\end{equation}
The linearized induction equations \eqref{eq:ind} are 
\begin{equation}
i\Delta\omega\delta B_r + ik_\phi B_\phi\delta v_r =0,
\end{equation}
\begin{equation}
i\Delta\omega\delta B_\phi + \bigg[\frac{\p v_\phi}{\p r}-\frac{v_\phi}{r}\bigg] \delta B_r - B_\phi\frac{\p\delta v_r}{\p r} = 0,
\end{equation}
\begin{equation}
i\Delta\omega\delta B_z - \frac{B_z}{r}\delta v_r - ik_\phi B_z\delta v_\phi + ik_\phi B_\phi\delta v_z - B_z\frac{\p\delta v_r}{\p r} =0.
\end{equation}
Lastly, the entropy conservation \eqref{eq:3} yields 
\begin{equation}
i\Delta\omega\frac{\delta P}{P} - i\Delta\omega\Gamma\frac{\delta \rho}{\rho} - \frac{\Gamma}{\mathrm{L_S}} \delta v_r =0.
\label{eq:22}
\end{equation}
The barotropic assumption gives $1/\mathrm{L_S}\rightarrow 0$, and eq \eqref{eq:22} simplifies to
\begin{equation}
\delta \Psi= c_s^2\frac{\delta\rho}{\rho}.
\label{eq:25}
\end{equation}
According to eq \eqref{eq:psi}, it is straightforward to see that the perturbed pressure is related to density by $\delta P=c_s^2\delta\rho$. 

\subsubsection{non-ideal MHD limit: pure $B_\phi$}

When non-ideal MHD effects apply, the continuity and entropy equations \eqref{eq:17} and \eqref{eq:25} remain unchanged, while the other equations are modified. We now express the linearized equations for pure $B_\phi$. The perturbed momentum equations are
\begin{align}
i\Delta\omega\delta v_r &+ 2\Omega\delta v_\phi - \frac{\p\delta\Psi}{\p r}  \nn \\ 
& + \frac{1}{4\pi\rho}\bigg[ik_\phi B_\phi\delta B_r - \frac{2B_\phi}{r} \delta B_\phi-B_\phi\frac{\p\delta B_\phi}{\p r}+\frac{B_\phi^2}{r}\frac{\delta\rho}{\rho} \bigg]\nn \\ 
&=0,
\end{align}
\begin{equation}
i\Delta\omega\delta v_\phi - \frac{\kappa^2}{2\Omega}\delta v_r - ik_\phi\delta \Psi+\frac{1}{4\pi\rho}\bigg[\frac{B_\phi}{r}\delta B_r\bigg]=0,
\end{equation}
\begin{equation}
i\Delta\omega\delta v_z + \frac{ik_\phi B_\phi}{4\pi\rho}\delta B_z=0.
\label{eq:37}
\end{equation}
The perturbed induction equations are 
\begin{align}
i\Delta\omega\delta B_r &+ ik_\phi B_\phi\delta v_r  \nn \\ 
&+[\eta_\mathrm{O}+\eta_\mathrm{A}]\bigg[\frac{\p^2\delta B_r}{\p r^2} + \frac{1}{r}\frac{\p\delta B_r}{\p r}-k_\phi^2\delta B_r-\frac{\delta B_r}{r^2}-\frac{2}{r}ik_\phi\delta B_\phi \bigg] \nn \\ 
& +\eta_\mathrm{H}[k_\phi^2 \delta B_z] \nn \\ 
&=0,
\end{align}
\begin{align}
i\Delta\omega\delta B_\phi &+ \bigg[\frac{\p v_\phi}{\p r}-\frac{v_\phi}{r}\bigg] \delta B_r - B_\phi\frac{\p\delta v_r}{\p r}  \nn \\ 
& +[\eta_\mathrm{O}+\eta_\mathrm{A}]\bigg[\frac{\p^2\delta B_\phi}{\p r^2} + \frac{1}{r}\frac{\p\delta B_\phi}{\p r}-k_\phi^2\delta B_\phi-\frac{\delta B_\phi}{r^2}+\frac{2}{r}ik_\phi\delta B_r \bigg] \nn \\ 
& +\eta_\mathrm{H}\bigg[ik_\phi\bigg(\frac{\p \delta B_z}{\p r} - \frac{\delta B_z}{r}\bigg) \bigg]\nn \\ 
&=0,
\end{align}
\begin{align}
i\Delta\omega\delta B_z &+ ik_\phi B_\phi\delta v_z\nn \\ 
& +\eta_\mathrm{O}\bigg[\frac{\p^2\delta B_z}{\p r^2} + \frac{1}{r}\frac{\p\delta B_z}{\p r}-k_\phi^2\delta B_z\bigg] \nn \\ 
& +\eta_\mathrm{A}\bigg[-k_\phi^2\delta B_z\bigg]  \nn \\ 
& +\eta_\mathrm{H}\bigg[ik_\phi\bigg(- \frac{\p \delta B_\phi}{\p r}-\frac{\delta B_\phi}{r} + ik_\phi\delta B_r\bigg)  \bigg]\nn \\ 
&=0.
\label{eq:40}
\end{align}
In the limit of pure $B_\phi$ and $k_z=0$, all three non-ideal MHD effects are present. 


\subsubsection{non-ideal MHD limit: pure $B_z$}

For a pure $B_z$ field, the vertical component of the momentum equation yields $\delta v_z =0$, while the remaining two linearized momentum equations are
\begin{align}
i\Delta\omega\delta v_r + 2\Omega\delta v_\phi - \frac{\p\delta\Psi}{\p r} -\frac{1}{4\pi\rho}\bigg[B_z\frac{\delta B_z}{\p r}\bigg]=0,
\end{align}
\begin{equation}
i\Delta\omega\delta v_\phi - \frac{\kappa^2}{2\Omega}\delta v_r - ik_\phi\delta \Psi + \frac{1}{4\pi\rho}\bigg[-ik_\phi B_z\delta B_z\bigg]=0.
\end{equation}
The non-ideal MHD effects manifest in the induction equation \eqref{eq:ind},
\begin{align}
i\Delta\omega\delta B_r & +\eta_\mathrm{O}\bigg[\frac{\p^2\delta B_r}{\p r^2} + \frac{1}{r}\frac{\p\delta B_r}{\p r}-k_\phi^2\delta B_r-\frac{\delta B_r}{r^2}-\frac{2}{r}ik_\phi\delta B_\phi \bigg] \nn \\ 
\label{eq:28}
&=0,
\end{align}
\begin{align}
i\Delta\omega\delta B_\phi & + \bigg[\frac{\p v_\phi}{\p r}-\frac{v_\phi}{r}\bigg] \delta B_r  \nn \\ 
& +\eta_\mathrm{O}\bigg[\frac{\p^2\delta B_\phi}{\p r^2} + \frac{1}{r}\frac{\p\delta B_\phi}{\p r}-k_\phi^2\delta B_\phi-\frac{\delta B_\phi}{r^2}+\frac{2}{r}ik_\phi\delta B_r \bigg] \nn \\ 
&=0,
\label{eq:29}
\end{align}
\begin{align}
i\Delta\omega\delta B_z &+B_z\bigg[\frac{1}{L_\rho}\delta v_r-\frac{i\Delta\omega}{c_s^2}\delta\Psi\bigg] \nn \\ 
& +[\eta_\mathrm{O}+\eta_\mathrm{A}]\bigg[\frac{\p^2\delta B_z}{\p r^2} + \frac{1}{r}\frac{\p\delta B_z}{\p r}-k_\phi^2\delta B_z\bigg] \nn \\ 
&=0.
\end{align}

In the limit of pure $B_z$ and $k_z=0$, only Ohmic resistivity and ambipolar diffusion contribute to the linearized induction equations, but not Hall drift. Ambipolar diffusion manifests only in the vertical component of the induction equation, exhibiting identical terms to Ohmic resistivity. 

\section{Numerical Methods}\label{sec:me}

We solve the linearized equations, which are ODE eigenvalue value problems, presented in \S\ref{sec:pertb} and Appendix \ref{app:kz} numerically. Two numerical methods are employed: the pseudospectral method (\S\ref{sec:sp}) and the finite difference method (\S\ref{sec:fd}). 

We find that generically the spectral method, via the Python package \textsc{dedalus}, outperforms the finite difference method. The finite difference method produces a number of spurious modes with oscillation frequencies $w_r$ that closely resembled those of RWI modes, particularly when incorporating Hall drift. Consequently, we use the spectral method as the primary numerical approach for solving the ODE systems. The details of these two methods are provided below.

\subsection{Pseudospectral method}\label{sec:sp}

Pseudospectral methods, also known as orthogonal collocation methods, approximate the solution of differential equations at selected collocation points, by a weighted sum of orthogonal basis functions, which are often chosen to be orthogonal polynomials up to a certain degree. 
Chebyshev polynomials of the first kind $T_n$, where $n=0,1,2,...,N-1$, are chosen in our problem, for which the eigenfunctions are expanded in. The radial domain, that spans $r\in[0.4,1.6]$, is discretized into $N$ Chebyshev collocation points. To minimize the interpolation error, these nodes are non-uniform and are selected as the roots of $N$th degree Chebyshev polynomial $T_N$, which cluster near the ends of the domain \citep[][Figure 1]{burns+20}. 

The differential equations described in \S\ref{sec:th} construct standard linear eigenvalue problems. Written compactly in a generalized matrix form, it is 
\begin{equation}
A\vec{x} = \mathcal{L}\vec{x} + \omega\mathcal{M}\vec{x} = 0,
\label{eq:}
\end{equation} 
where $\omega$ is the eigenvalue, $\vec{x} = [\vec{\delta v}_r,\vec{\delta B}_r,\vec{\delta\rho},...]^\mathrm{T}$ is a vector of eigenfunctions with $M$ perturbed quantities, and $A$, $\mathcal{L}$, $\mathcal{M}$ are $MN\times MN$ sized matrices, with $\mathcal{L}$ composed of linear operators. 
We employ \textsc{dedalus}\footnote{\url{https://dedalus-project.org/}}, a general purpose spectral code for differential equations \citep{burns+20}, to solve the linear eigenvalue problem described above. We utilize the dense solver method by \texttt{solve\_dense} in \textsc{dedalus}, for which matrix $A$ is converted to dense arrays, and the \texttt{scipy.linalg.eig} routine in Python is utilized to directly solve the eigenvalue problem.

The boundary condition employed is the density wave boundary condition, commonly used in the linear analysis of RWI \citep[e.g.,][]{li_etal00,ono_etal16}. Details of its implementation are provided in \S\ref{sec:fd}. In addition, we tested periodic boundary conditions with Fourier basis (\texttt{ComplexFourier} in \textsc{dedalus}) during the revision process. The Fourier basis significantly simplifies the computation of $k_r$ at the boundaries (\S\ref{sec:fd}), effectively capturing density waves at the domain ends while producing more robust results than density wave boundary condition. Thus, we recommend using Fourier basis for RWI linear analysis in the future work.

A numerical resolution of $N=256$ is adopted throughout. Spurious modes may arise due to the truncation of differential equations to finite-dimensional algebraic equations \citep{zebib87,Mcfadden+90}. Hence, we increase the numerical resolution to $N=400$ to filter out the spurious modes. By comparing solutions between low and high resolutions, we retain valid modes within a tolerance of $10^{-6}$, following an approach for mode rejection described in \textsc{Dedalus Eigentools
}\footnote{\url{https://eigentools.readthedocs.io/en/latest/}}. We note that the RWI modes computed at $N=256$ are consistent with $N=400$.

\subsection{Finite difference method}\label{sec:fd}

In finite difference method, we approximate the ODEs by finite difference equations, a technique previously adopted in RWI literature by \citet{ono_etal16,lb23}. We discretize the ODEs on a grid with $N=1001$ points and uniform spacing $h$, spanning the domain $r\in[0.4,1.6]$. The system can be compactly expressed using $MN\times MN$ sized matrix $A$, corresponding to the set of ODEs, is written as
\begin{equation}
A\vec{x} = 
\begin{vmatrix}
A_{00} & ... & A_{0,M-1}\\
... & ... & ... \\
A_{M-1,0} & ... & A_{M-1,M-1}
\end{vmatrix}\vec{x}=0.
\end{equation}
Matrix $A$ consists of $M\times M$ submatrices $A_{ij}$, for $i,j = 0,...,M-1$. The value of $M$ is set by the number of perturbed quantities contained in the system of ODEs. Each submatrix $A_{ij}$ has a size of $N\times N$. The vector of perturbed quantities is expressed as $\vec{x} = [\vec{\delta v}_r,\vec{\delta B}_r,\vec{\delta\rho},...]^{\mathrm{T}}$, where each perturbed quantity, for example, $\delta\vec{v}_r=\delta\vec{v}_r(r_k)$, is evaluated at grid points $r_k$, for $k=0,...,N-1$. 

The entries of $A_{ij}$ are filled in by constructing the linear operators of the first derivative $d/dr$ and the second derivative $d^2/dr^2$. We adopt the central differencing scheme to obtain the linear operators of the first derivative \citep{lb23},
\begin{equation}
A_{ij,n,n-1} = -\frac{1}{2h}, 
\end{equation}
\begin{equation}
A_{ij,n,n+1} = \frac{1}{2h}, 
\end{equation}
and the second derivative,
\begin{equation}
A_{ij,n,n-1} = \frac{1}{h^2}, 
\end{equation}
\begin{equation}
A_{ij,n,n} = -\frac{2}{h^2}, 
\end{equation}
\begin{equation}
A_{ij,n,n+1} = \frac{1}{h^2},
\end{equation}
for $n=1,...,N-2$.

The differential equations described in \S\ref{sec:th} form standard linear eigenvalue problems. The eigenvalue ($\omega$) and the corresponding eigenfunction (perturbed quantities $\vec{x}$) are readily determined by utilizing \texttt{numpy.linalg.eig} routine in Python.

To accomodate the density waves, WKBJ boundary conditions are imposed, and the linear operator $d/dr-ik_r=0$ is applied to the boundary.
In the finite difference method, the boundary conditions involve in the zeroth and $(N-1)$ th row of $A_{ij}$. Using the forward differencing scheme for the outer boundary conditions and the backward differencing scheme for the inner boundary conditions, the entries of $A_{ij}$ at boundaries can be written as
\begin{equation}
A_{ij,00} = -\frac{1}{h}, 
\end{equation}
\begin{equation}
A_{ij,01} = \frac{1}{h} - ik_r, 
\end{equation}
\begin{equation}
A_{ij,N-1,N-2} = -\frac{1}{h} - ik_r, 
\end{equation}
\begin{equation}
A_{ij,N-1,N-2} = \frac{1}{h}. 
\end{equation}

To solve for $k_r$ at the boundaries, away from the RWI region where the background quantities vary slowly over radius, we use WKBJ analysis by assuming $\delta v, \delta B, \delta\rho ... \propto\exp(ik_rr)$. Substituting the derivatives by $k_r$, we obtain $B(k_r,\omega)\vec{y}=0$ at the boundaries, where $B$ is a matrix containing $k_r$ and $\omega$, and $\vec{y} = [\delta v_r,\delta B_r,\delta\rho,...]^{\mathrm{T}}$. Both $B$ and $\vec{y}$ are evaluated at $r_0$ and $r_{N-1}$. We take $\omega$ to be the value obtained from last $\beta$. This approach works if we gradually decrease $\beta$ values in the domain of interest. Now the only unknown in matrix $B$ is $k_r$, which is determined using the Newton-Raphson method by setting the determinant of $B$ to zero, $\mathrm{det}\:B(k_r)=0$. The implementation of this method utilizes the source code of \texttt{scipy.optimize.newton} as a reference. The iteration converges within five to ten steps for a tolerance of $10^{-10}$.

We choose azimuthal mode number $m=2,3,4,5$ for the calculations presented in this work. For smaller or larger mode numbers, the growth rates can be relatively small and are sensitive to the grid resolution. We note that when $m=1$, the inner boundary condition differs from the WKB one described earlier. In the hydrodynamic limit, the inner boundary lies within the evanescent region, and the inner Lindblad resonance is absent. Instead, one can assume a power-law behavior in $r$ and ensure regularity at $r=0$ \citep{ono_etal16}.

\section{Results}\label{sec:re}

The numerical solutions are presented for the hydrodynamic limit (\S\ref{sec:hy}), ideal MHD limit (\S\ref{sec:ideal}), Ohmic resistivity (\S\ref{sec:ohm}), ambipolar diffusion (\S\ref{sec:ad}), and the Hall drift (\S\ref{sec:ha}), respectively. The RWI and MRI linear growth rates are compared in \S\ref{sec:mri}.
The analysis will be focused on the RWI growth rates, as the eigenfunctions are similar to those in the pure hydrodynamic limit; two Rossby waves are located on each side of the corotation radius, with density waves excited at the inner and outer Lindblad resonances, propagating away from the corotation radius \citep[e.g.,][]{lovelace+14}.
Note that in the Figures shown below, we select and present only the fastest growing modes, though a handful of slower growing modes also exist.

\subsection{Hydrodynamic limit}\label{sec:hy}

\begin{figure}
    \centering
    \includegraphics[width=0.5\textwidth]{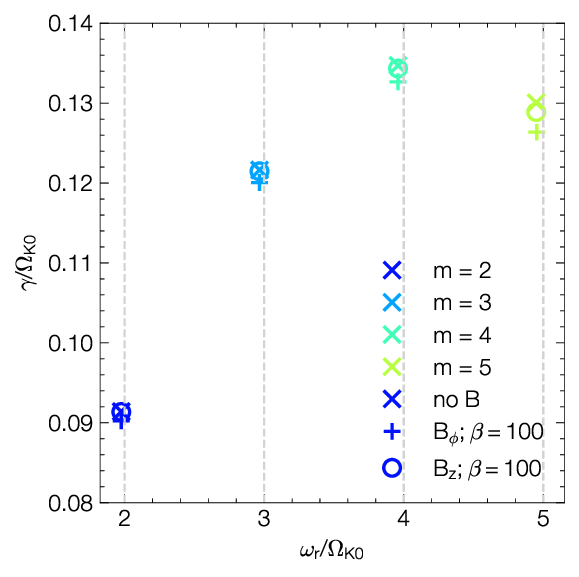}
    \caption{
    RWI growth rates $\gamma/\Omega_\mathrm{K0}$ vs oscillation frequencies $\omega_r/\Omega_\mathrm{K0}$ at $m=2,3,4,5$ in the pure hydrodynamic limit (crosses), pure $B_\phi$ ($\beta=100$; pluses), and pure $B_z$ ($\beta=100$; circles). 
    }
    \label{fig:hy}    
\end{figure}

We first revisit hydrodynamic results of RWI growth rates. Figure \ref{fig:hy} shows the growth rates $\gamma/\Omega_\mathrm{K0}$ versus oscillation frequencies $\omega_r/\Omega_\mathrm{K0}$ in the hydrodynamic limit (crosses). Specifically, they are $\omega/\Omega_\mathrm{K0} \approx 1.976+0.091i, 2.967+0.122i, 3.959 + 0.135i, 4.949 + 0.130i$ for $m=2,3,4,5$, respectively. It is clear that the growth rates peak at $m=4$. The oscillation frequencies $\omega_r-m\Omega_\mathrm{K0}\approx 0$ indicate that the Rossby waves are launched near the corotation radius \citep{lovelace+14}.
\citet{ono_etal16} performed a parameter study of the linear growth rates of the RWI for various background vortensity profiles. Their Figures 6 and 7 demonstrate that the most unstable azimuthal mode number $m$ depends on the amplitude and width of the profile. Our equilibrium disk, characterized by a Gaussian bump, closely mirrors their GB model, particularly in cases iii, iv, and v. By applying our parameters to their Figure 7, the most unstable modes are found to be $m = 4, 5$. This is consistent with Figure \ref{fig:hy}.
Moreover, on top of the hydrodynamic results, we plot the growth rates for pure $B_\phi$ (pluses) and pure $B_z$ (circles) fields at $\beta = 100$, representing a very weak magnetic field, hence approaching the hydrodynamic limit. Indeed, the growth rates at $\beta = 100$ show good agreement with the hydrodynamic solutions.

\begin{figure}
    \centering
    \includegraphics[width=0.5\textwidth]{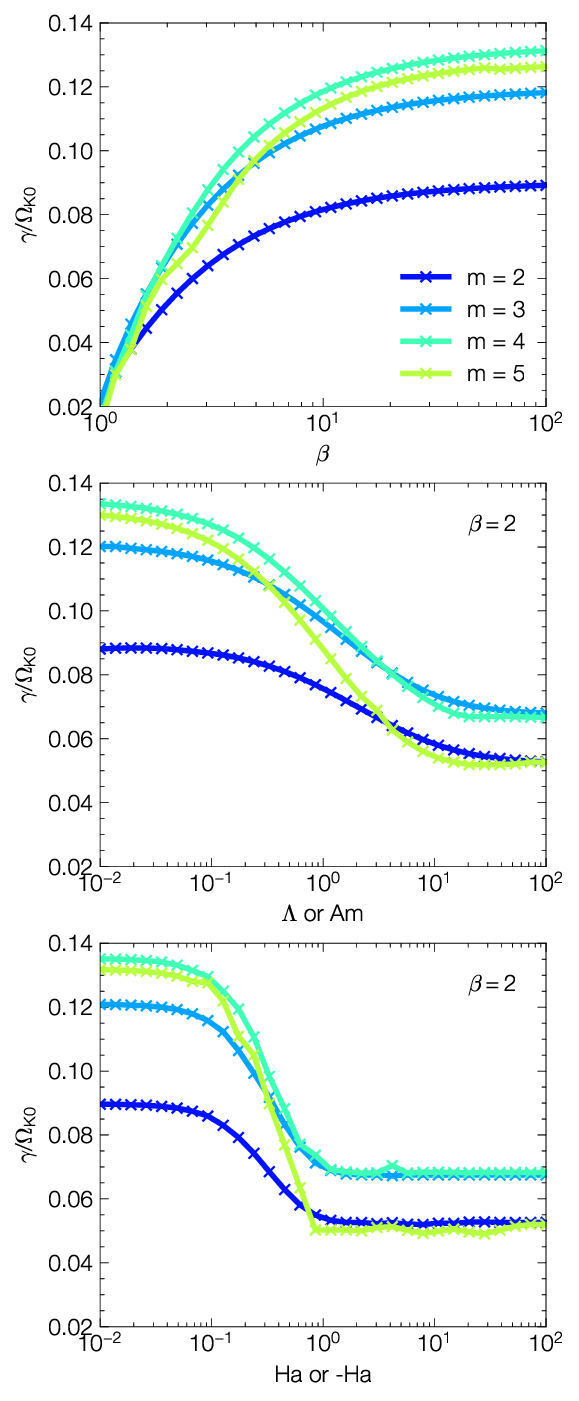}
    \caption{Pure $B_\phi$ and $k_z = 0$. 
    Top panel: RWI growth rates $\gamma/\Omega_\mathrm{K0}$ vs plasma $\beta$ in the ideal MHD limit for different $m$. 
    Middle panel: growth rates vs Ohmic ($\Lambda$) or ambipolar (Am) Els\"{a}sser numbers at $\beta=2$. 
    Bottom panel: growth rates vs Hall Els\"{a}sser number Ha or -Ha at $\beta=2$. 
    }
    \label{fig:bp}    
\end{figure}

\begin{figure}
    \centering
    \includegraphics[width=0.5\textwidth]{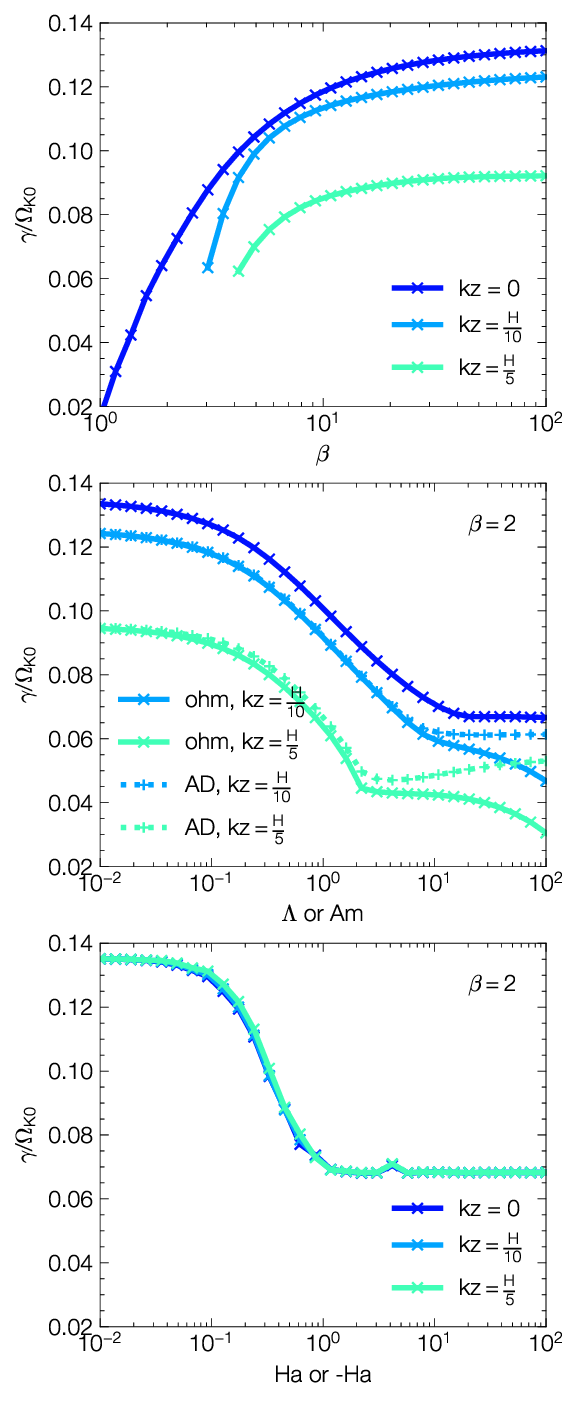}
    \caption{
    Pure $B_\phi$ and $k_z\neq 0$. 
    Top panel: RWI growth rates $\gamma/\Omega_\mathrm{K0}$ vs $\beta$ in the ideal MHD limit for different $k_z = 0, 1/10H, 1/5H$ at fixed $m=4$. 
    Middle panel: growth rates vs Ohmic ($\Lambda$; solid) or ambipolar (Am; dotted) Els\"{a}sser numbers at $\beta=2$. 
    Bottom panel: growth rates vs Hall Els\"{a}sser number Ha or -Ha at $\beta=2$. 
    }
    \label{fig:bp_kz}    
\end{figure}

\subsection{Ideal MHD}\label{sec:ideal}

we begin with the pure $B_\phi$ field to analyze the numerical solutions in the ideal MHD limit. The top panel of Figure \ref{fig:bp} depicts normalized RWI growth rates $\gamma/\Omega_\mathrm{K0}$ as a function of plasma $\beta$, for different azimuthal mode numbers $m=2,3,4,5$ at an infinite vertical wavelength, $k_z=0$. 
At $\beta=100$, the growth rates nearly recover the hydrodynamic RWI results, with the maximum growth rate observed at $m=4$ for the chosen set of parameters, as discussed in \S\ref{sec:hy}. As $\beta$ exceeds $100$, the curves begin to plateau. Conversely, as plasma $\beta$ decreases, the growth rates diminish for all azimuthal mode numbers $m$ investigated. This suggests that increased magnetization suppresses the RWI growth. 
It can be interpreted as that, in the ideal MHD limit, magnetic fields are perfectly coupled with the gas, potentially restricting the free motion of the perturbed fluid.
The top panel of Figure \ref{fig:bp_kz} shows growth rates for different vertical wavenumbers $k_z=0, 1/10H, 1/5H$ at $m=4$. It is evident that increasing vertical wavenumber tends to reduce RWI growth rates. 

Figure \ref{fig:bz} (top panel) shows the RWI growth rates in the ideal MHD limit for the pure $B_z$ field. Unlike pure $B_\phi$ field, where growth rates monotonically decrease towards smaller $\beta$, different azimuthal mode numbers $m$ exhibit distinct behaviors across the range of $\beta$ . The parameter space of $\beta$ is now extended down to $0.01$, in order to provide ample coverage to observe the trend in growth rates. For relatively large $m=5$, the RWI growth rates decrease monotonically as $\beta$ decreases, similar to pure $B_\phi$ field. Conversely, smaller $m=2,3,4$ show an increase in growth rates as $\beta$ decreases, followed by a decline as $\beta$ drops further. The peak of the growth rates occurs at higher $\beta$ values for larger $m$. This is in agreement with \citet{yu13}, who observed a peak in growth rate for $m=4$ around $\beta\approx 0.1$, as shown in their Figure 2. 
As $\beta$ approaches $0.01$, the growth rates asymptotically approach a stable value for each $m$. When there is a finite vertical wavelength ($k_z\neq 0$), the growth rates diminish with higher $k_z$, as depicted in the top panel of Figure \ref{fig:bz_kz}.

\subsection{Ohmic resistivity}\label{sec:ohm}

Next, we examine Ohmic resistive disks. The middle panel of Figure \ref{fig:bp} shows RWI growth rates as a function of the Ohmic Els\"{a}sser number $\Lambda$ at $\beta=2$. As $\Lambda\rightarrow\infty$, the growth rates tend to approach those of ideal MHD limit. For example, at $\beta\approx2.212$ the growth rates calculated for $m=2,3,4,5$ in the ideal MHD limit are $\omega/\Omega_\mathrm{K0} \approx 1.974+0.0554i,\ 2.963+0.0708i,\ 3.956+0.0726i,\ 4.949+0.0647i$, respectively (top panel; Figure \ref{fig:bp}). Correspondingly, the growth rates calculated for resistive disks at $\Lambda=100$ are $\omega/\Omega_\mathrm{K0}\approx 1.973+0.0528i,\  2.963+0.0666i,\ 3.956+0.0666i,\ 4.949+0.0526i$. 
As Ohmic Els\"{a}sser number decreases, the growth rates for all $m$ models gradually increase. At $\Lambda=10^{-2}$, they converge towards hydrodynamic results. The growth rates computed for the resistive disks at $\Lambda=0.01$ are $\omega/\Omega_\mathrm{K0}\approx 1.971+0.0881i,\  2.960+0.120i,\ 3.951+0.134i,\ 4.941+0.130i$, respectively, which are consistent with Figure \ref{fig:hy}.

\begin{figure}
    \centering
    \includegraphics[width=0.5\textwidth]{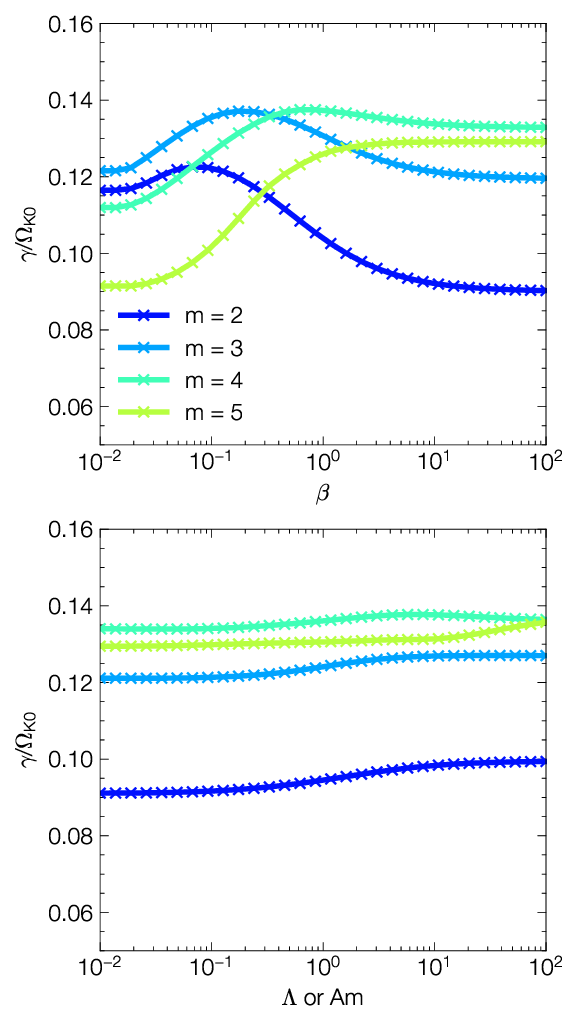}
    \caption{Same as top and middle panels of Figure 1, but for pure $B_z$ disks.}
    \label{fig:bz}    
\end{figure}

The pure $B_z$ disk follows the similar trend as the pure $B_\phi$ disk. As $\Lambda$ approaches infinity, the growth rates resemble those in the ideal MHD limit. Specifically, the growth rates at $\beta\approx2.212$ in the ideal MHD limit are $\omega/\Omega_\mathrm{K0}\approx 1.973+0.0979i, 2.963+0.126i, 3.956+0.136i, 4.949+0.128i$, respectively (top panel; Figure \ref{fig:bz}). For comparison, at $\beta=2$ and $\Lambda=100$, the RWI growth for $m=2,3,4,5$ are, respectively, $\omega/\Omega_\mathrm{K0} \approx 1.973+0.0994i, 2.964+0.127i, 3.960+0.136i, 4.956+0.136i$ (middle panel; Figure \ref{fig:bz}). 
As the Ohmic Els\"{a}sser number decreases, growth rates of different $m$ models all converge towards hydrodynamic results. The growth rates in the resistive disks at $\beta=2$ and  $\Lambda=0.01$ are $\omega/\Omega_\mathrm{K0}\approx 1.976+0.0911i,\  2.967+0.121i,\ 3.959+0.134i,\ 4.949+0.129i$, respectively. 

The middle panels of Figure \ref{fig:bp_kz} and Figure \ref{fig:bz_kz} show variations in growth rates with Ohmic Els\"{a}sser number $\Lambda$ at $m=4$ for different vertical wavenumbers $k_z$. It is evident that the growth rates decrease steadily with the vertical wavenumber $k_z$.

\subsection{Ambipolar diffusion}\label{sec:ad}

For $k_z=0$, ambipolar diffusion shares similar perturbation equations with Ohmic resistivity (\S\ref{sec:pertb}). Consequently, the curves for Am overlap with those for $\Lambda$ (middle panels, Figure \ref{fig:bp} and Figure \ref{fig:bz}).
For non-zero $k_z$, in the pure $B_\phi$ model, the growth rates in the ambipolar diffusion limit closely match those in the resistivity limit for $k_z=1/10H, 1/5H$, though with slight deviations observed when $\Lambda>2$ or Am$>2$ (middle panel, Figure \ref{fig:bp_kz}). 
However, for non-zero $k_z$ in the pure $B_z$ model, we suspect that the obtained modes may be spurious. Specifically, the RWI growth rates were found to increase indefinitely with $k_z$ for $\mathrm{Am} > 0.2$, and this behavior persists even at a resolution of $N=400$ using the spectral method. To ensure the reliability of our figures, we have omitted the curves for Am in the middle panel of Figure \ref{fig:bz_kz}. Future studies are needed to investigate and resolve this issue.

\subsection{The Hall drift}\label{sec:ha}

Unlike resistivity or ambipolar diffusion, Hall physics is sensitive to rotation. This is evident by reversing the sign of $B$ in the induction equation. In the limit of a vertical field, the Hall Els\"{a}sser number is written by $\mathrm{Ha}=v_{Az}^2/v_H^2$, where $v_H^2\equiv \Omega B_zc/(2\pi en_e)$ is the square of the Hall velocity \citep{bt01,lk22}. It follows that $v_H^2$ and Ha take negative (positive) values if $\bb{\Omega}$ and $\bb{B_z}$ are oriented oppositely (parallel). For a fixed direction of rotation, the toroidal field reverses its sign along with vertical field. We explore positive and negative Ha in pure $B_\phi$ and $B_z$ disk models.

In the case of pure $B_\phi$ and $k_z=0$, the Hall drift has a similar influence as resistivity and ambipolar diffusion (bottom panel; Figure \ref{fig:bp}). Moreover, the results obatined for Ha and -Ha overlap.
As Ha or -Ha$\rightarrow\infty$, the growth rate resembles those of ideal MHD, while as Ha or -Ha$\rightarrow 0$, the growth rates revive and approach hydrodynamic results. 
For example, at $\beta\approx 2$ for $m=2,3,4,5$ the growth rates in the ideal MHD limit are provided in \S\ref{sec:ohm}. In the Hall drift limit at $\mathrm{Ha}=100$, the corresponding growth rates are $\omega/\Omega_\mathrm{K0}\approx 1.973+0.0526i,\  2.963+0.0675i,\ 3.956+0.0683i,\ 4.933+0.0520i$, respectively. 
At $\beta=100$ and $m=2,3,4,5$, the ideal MHD growth rates are also given in \S\ref{sec:hy}. In the Hall drift limit at $\mathrm{Ha}=0.01$ and $\beta=2$, the corresponding growth rates are $\omega/\Omega_\mathrm{K0}\approx 1.971+0.0896i,\  2.960+0.121i,\ 3.950+0.135i,\ 4.949+0.133i$, respectively.
The bottom panel of Figure \ref{fig:bp_kz} shows growth rates at $k_z\neq 0$. The variation of vertical wavelengths does not significantly influence the RWI growth. The sign of Ha does not yield distinct values of $\gamma$ when $k_z\neq 0$ either.

\begin{figure}
    \centering
    \includegraphics[width=0.5\textwidth]{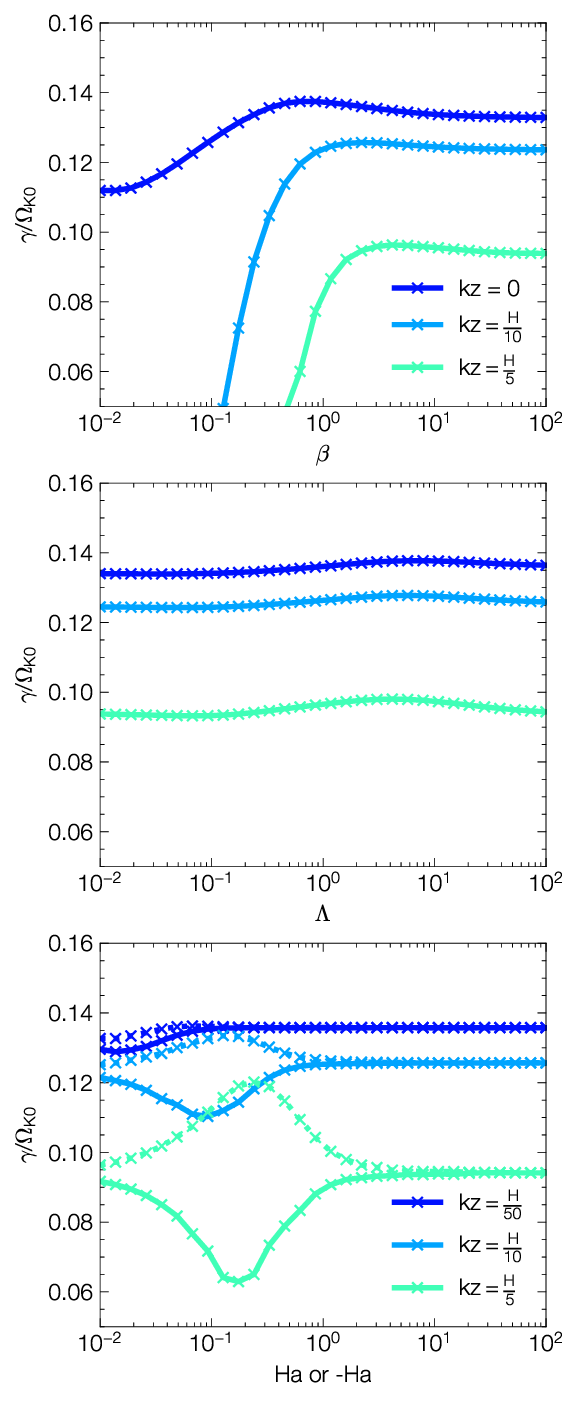}
    \caption{Same as Figure 2, but for pure $B_z$ disks. The middle panel only shows growth rates vs Ohmic Els\"{a}sser numbers ($\Lambda$).
    The bottom panel shows positive (solid) and negative (dotted) Ha at $k_z = 1/50H, 1/10H, 1/5H$, respectively.
    }
    \label{fig:bz_kz}    
\end{figure}

In the case of pure $B_z$, the Hall terms only appear in the induction equations when the wavelength is finite ($k_z\neq 0$). As shown in bottom panel of Figure \ref{fig:bz_kz}, when vertical wave number is very small, for example $k_z=1/50H$, the growth rates do not vary much across Ha or -Ha.
Increasing $k_z$ generically diminish the RWI growth. Furthermore, positive and negative Ha exhibit distinct behaviors; positive Ha results in a trough at $\mathrm{Ha}\approx 0.1$, whereas negative Ha results in a peak at around the same location.

\subsection{Comparison to MRI linear modes}\label{sec:mri}

We compare the linear growth between RWI and MRI. We follow the framework, derivations, and analyses described in \citet{lk22} for MRI. In most cases, we consider MRI channel modes with 
$k_x/k_z=0$, a pure vertical field $B_z$ and $\beta_z$, Keplerian rotation, and no vertical shear. We also consider azimuthal fields in the ambipolar diffusion dominated regime.

We start with the simplest ideal MHD limit. Eq (29) in \citet{lk22} shows the bi-quadratic dispersion relation, and the MRI growth rate $s$ is found to be
\begin{equation}
\frac{2s^2}{\Omega^2} = - \Bigg[\frac{2k^2_zv^2_{Az}}{\Omega^2}+1 \Bigg] + \Bigg[\frac{16k^2_zv^2_{Az}}{\Omega^2}+1 \Bigg]^{1/2},
\end{equation}
where $v_{Az}=B_z/\sqrt{4\pi\rho}$ is the vertical Alfv\'{e}n velocity.
The maximum MRI growth rate $s/\Omega=3/4$ occurs when $k^2_zv^2_{Az}=15/16\Omega^2$ as expected. On the other hand, growth rates vanish when $k^2_zv^2_{Az}=3\Omega^2$. 
In a realist protoplanetary disk, the vertical wavelength should be shorter than the pressure sclae height, $k_z>1/H$. Therefore, the criteria for the existence of MRI modes is $\beta_z>q_r^{-1}$, where $q_r=-\p\ln\Omega/\p\ln R$ is the dimensionless orbital shear, and has $q_r=3/2\sim O(1)$ for Keplerian rotation. The maximum growth can occur when $\beta_z>32/15$. Both criteria can be readily satisfied in protoplanetary disks, allowing linear MRI modes to easily surpass RWI modes.

For resistivity dominated disks, the instability criterion for MRI channel modes is presented in eq (34) of \citet{lk22}. Requiring the vertical wavelengths to be longer than the thickness of the disk yields
\begin{equation}
\beta_z > q_r^{-1}(1+\Lambda^{-2}).
\end{equation}
If $\Lambda\rightarrow\infty$, the system is almost in the ideal MHD regime, and $\beta_z>q_r^{-1}$ is required for channel modes to emerge. Thus, MRI surpasses RWI as shown above. 
If $\Lambda\ll 1$, the system is strongly resistivity dominated, and $\beta_z\gtrsim\Lambda^{-2}$. In this regime, RWI modes resemble hydrodynamic results, and can surpass MRI.
For a zero $B_\phi$ field, the instability criteria and physical behavior of ambipolar diffusion is very similar to the Ohmic regime. 

For a non-zero $B_\phi$ field, the MRI channel modes are diminished by small Am, for which regime ambipolar diffusion shear instability (ADSI) can emerge \citep{desch04,kb04,kunz08}. We start with the rather simple pure MRI channel modes. These modes exist when
\begin{equation}
\beta_z > q_r^{-1}(1+\mathrm{Am}^{-2}B^2/B_z^2).
\end{equation}
The presence of azimuthal fields contributes to stabilizing the MRI. In three-dimensional global numerical simulations incorporating ambipolar diffusion, a typical ratio of $B_\phi/B_z\sim O(10)$ has been found \citep{cb21}. Using this ratio, channel modes occur under conditions where $\beta_z \gtrsim 10^4, 10^2, 1$ for $\mathrm{Am}=0.1,1,10$, respectively. Thereby, RWI may dominate over MRI for $\mathrm{Am}\lesssim 0.1$ in a protoplanetary disk.
On the other hand, ADSI becomes significant when channel modes are diminished. ADSI modes have $|k_x/k_z|>0$. Although these modes can always emerge for sufficiently large $|k_x/k_z|$, their growth rates might be rather small \citep{lk22}. To investigate the dominance between MRI/ADSI and RWI, we compute the growth rates by the ambipolar  dispersion relation shown in eq (31) of \citet{kb04}. We utilize $B_\phi/B_z\sim O(10)$ and set a critical growth rate threshold at $0.01\OmK$. Under these conditions, no modes satisfy the growth rate criterion for $\mathrm{Am}<1$, leading us to conclude that RWI surpasses MRI/ADSI in this regime.

Lastly, we examine the Hall dominated regime. Positive Ha exhibits fast growth rates around $\sim 3/4\Omega$, stemming from a blend of MRI and Hall shear instability. There are three regimes associate with negative Ha. For $\mathrm{Ha}<-0.25$, we are in the Hall modified and diffusive MRI regimes, for which growth rates are rather fast $\sim 3/4\Omega$. For $\mathrm{Ha}>-0.25$, there is no instability possible, rendering it ideal for RWI modes to grow.

\section{Conclusions}\label{sec:c}

We studied the magnetized and weakly ionized RWI by Eulerian perturbations. The framework is three-dimensional and radially global. Ideal MHD and non-ideal MHD effects, including Ohmic resistivity, ambipolar diffusion, and the Hall drift are considered. The spectral method via \textsc{dedalus} is employed to resolve the eigenmodes. Our results are summarized as follows. 

For a pure $B_\phi$ field:
\begin{itemize}
\item 
When gas and magnetic fields are perfectly coupled, RWI growth rates increase with $\beta$. Strong magnetization tends to impede RWI. For all three non-ideal MHD effects, as Elsässer numbers approach infinity, the results resemble the ideal MHD limit; as Elsässer numbers approach zero, the results resemble the hydrodynamic limit. In the limits of ideal MHD, Ohmic resistivity, and ambipolar diffusion, non-zero vertical wavenumbers generically diminish RWI growth compared to $k_z=0$ limit. In Hall-dominated disks, non-zero wavenumbers do not significantly impact the results. The sign of Ha does not yield different results.
\end{itemize}

For a pure $B_z$ field:
\begin{itemize}
\item 
In the ideal MHD limit, RWI growth rates can either increase or decrease with $\beta$, depending on the azimuthal mode number $m$. In the limit of resistivity, similar to the $B_\phi$ model, as Elsässer numbers approach infinity, the results resemble the ideal MHD limit. As Elsässer numbers approach zero, the results resemble the hydrodynamic limit. In the limit of ideal MHD and resistivity, similar to the $B_\phi$ model, vertical wavenumbers generically diminish RWI growth. Hall drift only appears when $k_z\neq 0$. The sign of Ha slightly complicates the growth rates. 
\end{itemize}


\section*{Acknowledgements}

We thank the anonymous referee for the valuable comments that enhanced the clarity of this paper. CC expresses sincere gratitude for the support received from Professor Yanqin Wu at the University of Toronto. CC acknowledges funding from NSERC Canada and UK STFC grant ST/T00049X/1. AT acknowledges summer studentship from the Centre for Mathematical Sciences, University of Cambridge. CY is supported by the National SKA Program of China (grant 2022SKA0120101) and the National Natural Science Foundation of China (grants 11873103 and 12373071). MKL is supported by the National Science and Technology Council (grants 112-2112-M-001-064-, 113-2124-M-002-003-) and an Academia Sinica Career Development Award (AS-CDA-110-M06). 

\section*{Data Availability}

The data underlying this article will be shared on reasonable request to the corresponding author.


\bibliographystyle{mnras}
\bibliography{disk} 

\appendix
\onecolumn

\section{Linearized equations for $k_z\neq 0$}\label{app:kz}
\subsection{ideal MHD}

The linearized continuity equation is
\begin{equation}
\frac{i\Delta\omega}{c_s^2}\delta\Psi - \frac{\p\delta v_r}{\p r} - \bigg[\frac{1}{r}+\frac{1}{\mathrm{L_\rho}}\bigg]\delta v_r - ik_\phi\delta v_\phi - ik_z\delta v_z =0.
\end{equation}
The linearized momentum equations are
\begin{align}
i\Delta\omega\delta v_r &+ 2\Omega\delta v_\phi - \frac{\p\delta\Psi}{\p r} +\frac{1}{4\pi\rho}\bigg[i(k_\phi B_\phi + k_z B_z) \delta B_r-\frac{2B_\phi}{r}\delta B_\phi -B_z\frac{\delta B_z}{\p r} -B_\phi\frac{\delta B_\phi}{\p r}+\frac{B_\phi^2}{r}\frac{\delta\rho}{\rho}\bigg] =0,
\end{align}
\begin{align}
i\Delta\omega\delta v_\phi &- \frac{\kappa^2}{2\Omega}\delta v_r - ik_\phi\delta \Psi + \frac{1}{4\pi\rho}\bigg[\frac{B_\phi}{r}\delta B_r-ik_\phi B_z\delta B_z + ik_zB_z\delta B_\phi\bigg] = 0,
\end{align}
\begin{equation}
i\Delta\omega\delta v_z - ik_z\delta\Psi + \frac{1}{4\pi\rho}\bigg[ ik_\phi B_\phi\delta B_z - ik_zB_\phi\delta B_\phi  \bigg] =0,
\end{equation}
The linearized induction equations are 
\begin{equation}
i\Delta\omega\delta B_r + i(k_\phi B_\phi+k_zB_z)\delta v_r =0,
\end{equation}
\begin{equation}
i\Delta\omega\delta B_\phi + \bigg[\frac{\p v_\phi}{\p r}-\frac{v_\phi}{r}\bigg] \delta B_r - B_\phi\frac{\p\delta v_r}{\p r} + ik_z[B_z \delta v_\phi-B_\phi\delta v_z] = 0,
\end{equation}
\begin{equation}
i\Delta\omega\delta B_z - \frac{B_z}{r}\delta v_r - ik_\phi B_z\delta v_\phi + ik_\phi B_\phi\delta v_z - B_z\frac{\p\delta v_r}{\p r} =0.
\end{equation}

\subsection{non-ideal MHD limit: pure $B_\phi$}

The non-ideal MHD effects manifest in the induction equations,
\begin{align}
i\Delta\omega\delta B_r &+ ik_\phi B_\phi\delta v_r  \nn \\ 
&+\eta_\mathrm{O}\bigg[\frac{\p^2\delta B_r}{\p r^2} + \frac{1}{r}\frac{\p\delta B_r}{\p r}-(k_\phi^2+k_z^2)\delta B_r-\frac{\delta B_r}{r^2}-\frac{2}{r}ik_\phi\delta B_\phi \bigg] \nn \\ 
&+\eta_\mathrm{A}\bigg[ -\frac{\delta B_\phi}{r} -\frac{\p \delta B_\phi}{\p r}  + ik_\phi\delta B_r \bigg]ik_\phi \nn \\ 
& +\eta_\mathrm{H}\bigg[- ik_\phi\delta B_z + ik_z\delta B_\phi \bigg]ik_\phi \nn \\ 
&=0,
\end{align}
\begin{align}
i\Delta\omega\delta B_\phi & + \bigg[\frac{\p v_\phi}{\p r}-\frac{v_\phi}{r}\bigg] \delta B_r - B_\phi\frac{\p\delta v_r}{\p r} - ik_zB_\phi\delta v_z  \nn \\ 
& + [\eta_\mathrm{O}+\eta_\mathrm{A}]\bigg[\frac{\p^2\delta B_\phi}{\p r^2} + \frac{1}{r}\frac{\p\delta B_\phi}{\p r}-(k_\phi^2+k_z^2)\delta B_\phi-\frac{\delta B_\phi}{r^2}+\frac{2}{r}ik_\phi\delta B_r \bigg] \nn \\ 
& +\eta_\mathrm{H}\bigg[(- ik_z\delta B_r + \frac{\p \delta B_z}{\p r})ik_\phi + (- ik_\phi\delta B_z + ik_z\delta B_\phi) \frac{1}{r}\bigg]  \nn \\ 
&=0,
\end{align}
\begin{align}
i\Delta\omega\delta B_z &+ ik_\phi B_\phi\delta v_z\nn \\ 
& +\eta_\mathrm{O}\bigg[\frac{\p^2\delta B_z}{\p r^2} + \frac{1}{r}\frac{\p\delta B_z}{\p r}-(k_\phi^2+k_z^2)\delta B_z\bigg] \nn \\ 
& +\eta_\mathrm{A}\bigg[-k_\phi^2\delta B_z + k_\phi k_z\delta B_\phi \bigg]  \nn \\ 
& +\eta_\mathrm{H}\bigg[ -\frac{\p\delta B_\phi}{\p r} - \frac{\delta B_\phi}{r} + ik_\phi\delta B_r \bigg]ik_\phi  \nn \\ 
&=0.
\end{align}

\subsection{non-ideal MHD limit: pure $B_z$}\label{sec:}

The linearized induction equations are written as
\begin{align}
i\Delta\omega \delta B_r & + ik_z B_z \delta v_r \nn \\ 
& +\eta_\mathrm{O}\bigg[\frac{\p^2\delta B_r}{\p r^2} + \frac{1}{r}\frac{\p\delta B_r}{\p r}-(k_\phi^2+k_z^2)\delta B_r-\frac{\delta B_r}{r^2}-\frac{2}{r}ik_\phi\delta B_\phi \bigg] \nn \\ 
& +\eta_\mathrm{A}\bigg[-k_z^2\delta B_r - ik_z\frac{\p\delta B_z}{\p r} \bigg] \nn \\ 
& +\eta_\mathrm{H}\bigg[-ik_\phi\delta B_z + ik_z\delta B_\phi \bigg]ik_z \nn \\ 
&=0,
\end{align}
\begin{align}
i\Delta\omega\delta B_\phi &+ \bigg[\frac{\p v_\phi}{\p r}-\frac{v_\phi}{r}\bigg] \delta B_r  + ik_zB_z \delta v_\phi \nn \\ 
& +\eta_\mathrm{O}\bigg[\frac{\p^2\delta B_\phi}{\p r^2} + \frac{1}{r}\frac{\p\delta B_\phi}{\p r}-(k_\phi^2+k_z^2)\delta B_\phi-\frac{\delta B_\phi}{r^2}+\frac{2}{r}ik_\phi\delta B_r \bigg] \nn \\ 
& +\eta_\mathrm{A}\bigg[k_zk_\phi\delta B_z - k_z^2\delta B_\phi  \bigg]\nn \\ 
& +\eta_\mathrm{H}\bigg[- ik_z\delta B_r + \frac{\p \delta B_z}{\p r}\bigg]ik_z \nn \\ 
&=0,
\end{align}
\begin{align}
i\Delta\omega\delta B_z &+B_z\bigg[\frac{1}{L_\rho}\delta v_r-\frac{i\Delta\omega}{c_s^2}\delta\Psi\bigg] \nn \\ 
& +[\eta_\mathrm{O}+\eta_\mathrm{A}]\bigg[\frac{\p^2\delta B_z}{\p r^2} + \frac{1}{r}\frac{\p\delta B_z}{\p r}-(k_\phi^2+k_z^2)\delta B_z\bigg] \nn \\ 
& +\eta_\mathrm{H}\bigg[-\frac{\p\delta B_\phi}{\p r}- \frac{\delta B_\phi}{r} + ik_\phi\delta B_r \bigg]ik_z  \nn \\ 
&=0.
\end{align}

\bsp	
\label{lastpage}
\end{document}